\documentclass[a4paper]{jpconf}
\usepackage{amsfonts}
\begin{document}

\newcommand{\be}{\begin{equation}}
\newcommand{\ee}{\end{equation}}
\newcommand{\bea}{\begin{eqnarray}}
\newcommand{\eea}{\end{eqnarray}}
\newcommand{\beann}{\begin{eqnarray*}}
\newcommand{\eeann}{\end{eqnarray*}}
\newcommand{\nn}{\nonumber}

\newcommand{\Ref}[1]{(\ref{#1})}

\def\eps{\epsilon}
\def\tr{\mathop{\rm Tr}\nolimits}
\def\arctanh{\mathop{\rm arctanh}\nolimits}
\def\arctan{\mathop{\rm arctan}\nolimits}
\def\ln{\mathop{\rm ln}\nolimits}

\title{Scalar-Tensor theory as a singular subsector of $\Lambda(\phi)$ Plebanski gravity}

\author{David Beke}

\address{Faculty of Applied Sciences EA16, Gent University, Galglaan 2, 9000 Gent, Belgium}

\ead{david.beke@ugent.be}

\begin{abstract}
It is shown that, in the absence of matter fields, the coupling of a scalar field to the non-chiral Plebanski action can be obtained by relaxing the trace component of the simplicity constraints. This is realized by considering a subclass of generalized theories, where a potential depending on invariants of the Lagrange multipliers is added to the Plebanski action. Generically such theories propagate eight degrees of freedom, but here the (singular) subclass with the potential only depending on the trace of the Lagrange multiplier is displayed as the Bergmann-Wagoner-Nordtvedt class of scalar-tensor theories. 
 
\end{abstract}

\section{Introduction}

In the non-chiral Plebanski formalism, general relativity is described by a set of Lorentz-algebra-valued bivectors $B^{IJ}_{\mu\nu}$. These bivectors are put into correspondence with a tetrad by imposing the constraints 
\be
B^{IJ}\wedge B^{KL}=-\frac{1}{24}\eps^{IJKL}\eps_{MNPQ}B^{MN}\wedge B^{PQ},\label{SimplCons}
\ee
which are solved by
\bea
B^{IJ} & = & \pm\left(\star e\wedge e\right)^{IJ}, \label{Bsol}\\
B^{IJ} & = & \pm e^I\wedge e^J \label{Btop},
\eea
where $\star$ is the hodge dual in the Lie algebra. The correct dynamics is obtained by taking the kinetic part of the action to be BF:
\be
S_{Pl}\left[B^{IJ}_{\mu\nu},A_\mu^{IJ},\phi_{IJKL}\right] = \int B^{IJ}\wedge F_{IJ}(A)-\frac{1}{2}\left(\phi_{IJKL}+\frac{\Lambda}{6}\eps_{IJKL}\right)B^{IJ}\wedge B^{KL}.\label{Pleb}
\ee
Here, the Lagrange multiplier $\phi^{IJKL}$ is subject to $\eps_{IJKL}\phi^{IJKL}=0$, such that variation wrt it yields \Ref{SimplCons}. One has to impose reality conditions on $B^{IJ}$ in order to guarantee that $e^I_\mu e^K_\nu\eta_{IJ}$ is a real Lorentzian metric. We switch to the Euclidian theory by taking $B$ to be real and $\mathfrak{so}(4)$-valued, in order to bypass this technicality, but if the usual reality conditions are imposed, the conclusions hold for the Lorentzian theory. 

A class of modifications, where the constraints \Ref{SimplCons} are relaxed by promoting $\Lambda$ to a function of the scalar invariants of $\phi_{IJKL}$,
\be
S _{\Lambda(\phi)}=  \int B^{IJ}\wedge F_{IJ}- \frac{1}{2}\left(\phi_{IJKL}+\frac{\Lambda(\phi)}{6}\eps_{IJKL}\right)B^{IJ}\wedge B^{KL}\label{ModifPleb},
\ee
has been studied in the literature\cite{Alexandrov2009, Speziale2010}. The conclusion is that, generically, such theories are plagued by a very basic problem, containing a massless graviton and a massive graviton, coupled to the Boulware-Deser ghost. Here, a singular subclass of these theories is presented, where the dynamical degrees of freedom (DOFs) are a massless graviton and a scalar field. A more detailed discussion will be given in \cite{BS}.

\section{Irreducible decomposition of the simplicity constraints}

The $\mathfrak{so}(4)$-valued bivectors $B^{IJ}_{\mu\nu}$ can be parametrized in the following way \cite{Krasnov2008,Freidel2008}:
\be
B^{IJ}=\eta P^{IJ}_{(+),i}b^i_j\Sigma^j_{(+)\mu\nu}[e^I_\mu]+\bar{\eta}P^{IJ}_{(-),i}\bar{b}^i_j\Sigma^i_{(-)\mu\nu}[\bar{e}^I_{\mu}],\label{BDecomp}
\ee
where we introduced two independent tetrads $e^I_\mu$ and $\bar{e}^I_\mu$ and two independent unimodular fields $b^i_j$ and $\bar{b}^i_j$, and $\eta=\pm1$, $\bar{\eta}=\pm1$. $\Sigma^i=e^0\wedge e^i+\frac 1 2 \eps^i{}_{jk}e^j\wedge e^k$ is a basis for the space of self-dual, wrt $e^I_\mu e^J_\nu\eta_{IJ}$, two-forms. \footnote{One can check that both a rotation and a boost of the tetrad induce a rotation of $\Sigma^i$. The 36 components of $B^{IJ}_{\mu\nu}$ are thus decomposed in the $2\times 16$ components of two tetrads and the $2\times 8$ components of the unimodular fields minus $2\times (3+3)$ components because of rotational invariance in the $\mathfrak{su}(2)$ contractions - and the fact that $3+3$ components of $e^I_\mu$ label such a rotation.}

Equation \Ref{SimplCons} can be decomposed in its irreducibles, $({\mathbf 2},{\mathbf 0})\oplus({\mathbf 0},{\mathbf 2})\oplus({\mathbf 1},{\mathbf 1})\oplus ({\mathbf 0},{\mathbf 0})$, yielding
\bea
&&2P_{\bf(2,0)}^{IJKL}{}_{ij}e\left(m^{ij}-\frac1 3 m\delta^{ij}\right) + 2P_{\bf(0,2)}^{IJKL}{}_{ij}\bar{e}\left(\bar{m}^{ij}-\frac1 3\bar{m}\delta^{ij}\right)\label{ConsDec}\\ 
&&\quad\quad\quad\quad\quad\quad+ P_{\bf(1,1)}^{IJKL}{}_{ij}b^i_k\bar{b}^j_l\Sigma^k[e^I_\mu]\wedge\bar{\Sigma}[\bar{e}^I_\mu] + 2P_{\bf(0,0)}^{IJKL} \left(em-\bar{e}\bar{m}\nn\right)= 0,
\eea
where $m^{ij}=b^{i}_kb^{jk}$, $e=\det(e^I_\mu)$ and $m=m^i{}_i$.
The $\bf(2,0)$, resp. $\bf(0,2)$, components imply, due to the unimodularity, $m^{ij}=\delta^{ij}$, resp. $\bar{m}^{ij}=\delta^{ij}$. The solutions of the $\bf(1,1)$ components are  given by $\bar{e}^I_\mu\propto e^I_\mu$. From the scalar component it then follows that $\bar{e}^I_\mu=e^I_\mu$,  such that
\be
B^{IJ}_{\mu\nu}=\frac{\eta+\bar{\eta}}{2}e^I\wedge e^J +\frac{\eta-\bar{\eta}}{2}   (\star e\wedge e)^{IJ}
\ee
For $\eta=\bar{\eta}$, one obtains \Ref{Btop} which upon substitution in (\ref{Pleb}) results in $S=\pm\int e^I\wedge e^J\wedge F_{IJ}[A]$ which is a topological theory. For $\eta=-\bar{\eta}$, solution (\ref{Bsol}) and consequently the Einstein-Cartan (EC) action is recovered.



\section{Massive Scalar Tensor theory from modified constraints}

A `weak' relaxation of the constraints is to drop only one: the $\bf(0,0)$ component of (\ref{ConsDec}). In that case $\bar{e}^I_{\mu}$ and $e^I_{\mu}$ are not equal, but are conformally related to one another:
\be
b^{i}_j=\bar{b}^{i}_j=\delta^{i}_j, \quad \bar{e}^I_{\mu}=\psi e^I_{\mu}, \label{ConstrSols}
\ee
such that
\be
B^{IJ}=\frac{\eta+\bar{\eta} \psi^2}{2} \left(\star e\wedge e\right)^{IJ} + \frac{\eta- \bar{\eta} \psi^2}{2} e^I\wedge e^J. \label{BScalTetr}
\ee
Already at this level, one can expect to be dealing with a scalar-tensor theory, where the scalar degree of freedom results from the conformal factor relating the two metrics. 

In the class of theories described by (\ref{ModifPleb}), this relaxation is achieved by choosing
\be
\Lambda(\phi)=\Lambda_0(\tr(\phi)).\label{ScalarPotential}
\ee
As $\Lambda^{(1)}_{IJKL}=\frac{\delta \Lambda}{\delta \phi^{IJKL}}$ can only be solved for $\tr(\phi)$, this theory is not subject to the conclusions of \cite{Alexandrov2009} where it was shown that if $\Lambda^{(1)}$ can be solved for all components, \Ref{ModifPleb} contains eight DOFs. 

Varying the action (\ref{ModifPleb}) with $\Lambda$ of the form (\ref{ScalarPotential}) wrt $\phi_{IJKL}$ and taking the $\bf(2,0)$, $\bf(0,2)$ and $\bf (1,1)$ components indeed yields (\ref{ConstrSols}). The $\bf(0,0)$ component implies
\be
\Lambda'_0(\tr(\phi)) = \frac{-\delta_{IJKL}B^{IJ}\wedge B^{KL}}{\eps_{MNPQ}B^{MN}\wedge B^{PQ}} = -\frac{1-\psi^4}{2(1+\psi^4)}\label{TrEq}
\ee
which can be solved for $\tr(\phi)=\phi(\psi)$ (there are multiple sectors if this solution is not unique). 

The equations resulting from varying \Ref{ModifPleb} wrt the connection can be solved for the connection, and upon substitution in the action one obtains, up to a total derivative,
\be
\int e\left[\frac{\eta-\bar{\eta} \psi^2}{2}R-3\bar{\eta}\nabla_{\kappa}\psi\nabla^{\kappa}\psi- V_0(\psi)\left(1+\psi^4\right)\right]\label{EffAct}
\ee
where 
\be
V_0(\psi)=\Lambda_0(\phi(\psi)))+\frac{1}{2}\phi(\psi)\frac{1-\psi^4}{1+\psi^4}:
\ee
while the choice of $\Lambda_0$ does affect the potential terms, the kinetic terms are independent of it. 

One can diagonalize the kinetic terms by making a conformal field redefinition of the metric $\hat{g}_{\mu\nu}=|\frac{\eta-\bar{\eta}\psi^2}{2}|g_{\mu\nu}$, such that the action is this of a scalar field minimally coupled to this metric. One obtains, up to total derivatives,
\be
\int e \left[\pm R\mp \frac{6\eta\bar{\eta}}{(\eta-\bar{\eta}\psi^2)^2}\nabla_{\mu}\psi\nabla^{\mu}\psi-4\frac{1+\psi^4}{(\eta-\bar{\eta}\psi^2)^2}V_0(\psi)\right],
\ee
where we have dropped the hats, and where the + sign corresponds to $\eta-\bar{\eta}\psi^2>0$ and the - sign to $\eta-\bar{\eta}\psi^2<0$.  Note that all sectors have three DOFs, both for $\eta=-\bar{\eta}$ and $\eta=\bar{\eta}$, which would correspond to the topological sector of \Ref{Pleb}.
The sector with $\eta=-\bar{\eta}$, having the wrong sign of the kinetic term, is unstable. The signs of the kinetic terms in the other sectors do not pose problems, and for a suitable choice of the potential these are not plagued by instabilities.


\section{Coupling to matter} \label{sec:MC}

In order to extract physical predictions, one needs to understand matter coupling to $\Lambda(\phi)$ Plebanski gravity. Consider matter fields $Q$ for which the minimal coupling to the Einstein-Hilbert action is given by $S[g_{\mu\nu},Q]$. In the Plebanski case, coupling should reduce to the minimal coupling to EC theory, which is achieved by adding terms in the following way:
\be
S[B,A,\phi,Q]=S_{\rm{Pleb}}[B,A,\phi]+\displaystyle\sum_{i}\alpha_iS_m[G^{(i)}_{\mu\nu}(B^{IJ}_{\mu\nu}),Q],\label{TennieCouplingParam}
\ee
where $G^{(i)}$ is any metric constructed from $B^{IJ}_{\mu\nu}$, such that $G^{(i)}_{\mu\nu}(\star e \wedge e)=e^I_\mu e^J_\nu\eta_{IJ}$. After solving the simplicity constraints, this action indeed yields the EC action with the matter fields minimally coupled to it, such that at the classical level $\displaystyle\sum_i\alpha_i$ is the only genuine parameter. \footnote{
This coupling is a, for the Plebanski formulation irrelevant, generalization of the one discussed in \cite{Tennie2010}, where $\alpha_1=\alpha_2=\frac 1 2$ and $G^{(1)}_{\mu\nu}=e^I_\mu e^J_\nu\eta_{IJ}$, $G^{(2)}_{\mu\nu}=\bar{e}^I_\mu \bar{e}^J_\nu\eta_{IJ}$, with $e^I_\mu$ and $e^J_\nu$ as in \Ref{BDecomp}.
}

A natural generalization of (\ref{TennieCouplingParam}) to (\ref{ModifPleb}) is given by
\be
S_{\Lambda(\phi)}+\displaystyle\sum_{i}\alpha_iS_m[G^{(i)}_{\mu\nu}(g^{U+},g^{U-}),Q].\label{ModPlebCoupl}
\ee
One should not expect to find a unique way of coupling matter to a theory with 2 metrics. The parameters $\alpha_i, G^{(i)}$ should rather be viewed as genuine physical parameters. The weak equivalence principle can still be guaranteed, by requiring that all matter fields have the same parameters $\alpha_i,G^{(i)}$, such that the interaction of a test particle with the gravitational field is the same for all test particles.

Let us consider the physical interpretation of such coupling for the $\Lambda(\tr(\phi))$ subclass. Here the freedom in $G^{(i)}_{\mu\nu}(B^{IJ}_{\mu\nu})$ reduces to $f^{(i)}(\psi,\eta,\bar{\eta})g_{\mu\nu}$ with $f^{(i)}(1,\eta,-\eta)=1$.
Solving the modified simplicity constraints and the compatibility condition as in (\ref{EffAct}), one obtains upon substitution
\be
\int e\left[\frac{\eta-\bar{\eta} \psi^2}{2}R-3\bar{\eta}\nabla_{\kappa}\psi\nabla^{\kappa}\psi- V_0(\psi)\left(1+\psi^4\right)\right] + \displaystyle\sum_i \alpha_iS_m[f^{(i)}(\psi)g,Q].
\ee
Making once again the conformal field redefinition $\hat{g}_{\mu\nu}=|\frac{\eta-\bar{\eta}\psi^2}{2}|g_{\mu\nu}$ one obtains, after dropping the hats,
\be
\int e\left[\pm R\mp \frac{6\eta\bar{\eta}}{(\eta-\bar{\eta}\psi^2)^2}\nabla_{\kappa}\psi\nabla^{\kappa}\psi- 4\frac{1+\psi^4}{(\eta-\bar{\eta}\psi^2)^2}V_0(\psi)\right] + \displaystyle\sum \alpha_iS_m[\pm 2\frac{f^{(i)}(\psi)}{\eta-\bar{\eta}\psi^2}g,Q].
\ee
Of particular interest are couplings where only one term is present in the sum of matter terms: $\alpha_i=0$ when $i>1$. Generically, this yields a scalar-tensor theory, where $\psi$ is non-minimally coupled to the metric wrt which test particles move on geodesics.  A special case occurs for $f(\psi)=\frac{\eta-\bar{\eta}\psi^2}{2}$: $\psi$ is now minimally coupled to the physical metric. Since the matter fields do not interact directly with $\psi$, this corresponds to a dark matter field. In \cite{Speziale2010} it was shown that for a generic potential, $\frac{\eta}{2}g_{\mu\nu}-\frac{\bar{\eta}}{2}\bar{g}_{\mu\nu}$ contains precisely the massless graviton DOFs.  In other words, coupling to $\frac{\eta}{2}g_{\mu\nu}-\frac{\bar{\eta}}{2}\bar{g}_{\mu\nu}$ appears to be favored in a more general setting than the one discussed here.

\section{Conclusion}

It has been discussed that $\Lambda(\phi)$ Plebanski gravity, which generically has an unstable bi-metric nature, contains a subclass where the two metrics are conformally related to one another. This class has been shown to be equivalent to a subclass of the Bergmann-Wagoner-Nordtvedt scalar-tensor theories. As a technical result, this tells that if one is interested in gravity coupled to just a scalar field, this coupling does not need to be done by hand: it suffices to relax the $\bf{(0,0)}$ components of the constraints. If one could show how this scalar-tensor theory arises as an effective action, thereby giving a quantum origin to inflation\cite{La1989}, one could even draw physical conclusions from this result.

\section*{Acknowledgements}
We thank S. Speziale for cooperation in the work summarized here, and N. Van den Bergh for many stimulating discussions. The author is supported by the Research Foundation-Flanders (FWO).

\section*{References}

\bibliography{bibliog.bib}

\providecommand{\newblock}{}
\begin{thebibliography}{1}
\expandafter\ifx\csname url\endcsname\relax
  \def\url#1{{\tt #1}}\fi
\expandafter\ifx\csname urlprefix\endcsname\relax\def\urlprefix{URL }\fi
\providecommand{\eprint}[2][]{\url{#2}}

\bibitem{Alexandrov2009}
Alexandrov S and Krasnov K 2009 {\em Class. Quant. Grav.\/} {\bf 26} 055005
  {a}rXiv:0809.4763 [gr-qc]

\bibitem{Speziale2010}
Speziale S 2010 {\em Phys. Rev. D\/} {\bf 82} 064003 {a}rXiv:1003.4701 [hep-th]

\bibitem{BS}
Beke D and Speziale S {\em in preparation\/}

\bibitem{Krasnov2008}
Krasnov K 2008 {\em Class. Quant. Grav.\/} {\bf 25} 025001
  {a}rXiv:gr-qc/0703002

\bibitem{Freidel2008}
Freidel L 2008  {a}rXiv:0812.3200 [gr-qc]

\bibitem{Tennie2010}
Tennie F and Wohlfarth M 2010 {\em Phys. Rev. D\/} {\bf 82} 104052
  {a}rXiv:1009.5595 [gr-qc]

\bibitem{La1989}
La D and Steinhardt P 1989 {\em Phys. Rev. Lett.\/} {\bf 62}(4) 376--378

\end{thebibliography}

\end{document}